\documentclass[aps,prd,amsfonts,amssymb,preprint, eqsecnum,nofootinbib]
{revtex4} 
\usepackage{graphics,epsfig}
\begin{document}
\preprint{WM-04-107}
%
\title{\vspace*{0.5in} Higgsless GUT Breaking and Trinification \vskip
0.1in}
\author{Christopher D. Carone}\email[]{carone@physics.wm.edu}
\author{Justin M. Conroy}\email[]{jmconr@wm.edu}
\affiliation{Particle Theory Group, Department of Physics,
College of William and Mary, Williamsburg, VA 23187-8795}
\date{June 2004}
\begin{abstract}
Boundary conditions on an extra-dimensional interval can be chosen to
break bulk gauge symmetries and to reduce the rank of the gauge group.
We consider this mechanism in models with gauge trinification.  We
determine the boundary conditions necessary to break the trinified
gauge group directly down to that of the standard model.  Working in an
effective theory for the gauge symmetry-breaking parameters on a
boundary, we examine the limit in which the GUT-breaking sector is
Higgsless and show how one may obtain the low-energy particle content
of the minimal supersymmetric standard model.  We find that gauge unification
is preserved in this scenario, and that the differential gauge coupling
running is logarithmic above the scale of compactification.  We compare
the phenomenology of our model to that of four-dimensional trinified
theories.
\end{abstract}
\maketitle

\section{Introduction}\label{sec:intro}
Extra spatial dimensions allow for the possibility of gauge symmetry
breaking by the appropriate choice of boundary conditions on the
fields.  The relevance of this point to model building was first
realized by Kawamura~\cite{kawamura}, in the context of SU(5) grand
unified theories (GUTS), and was developed substantially afterwards by
a number of authors~\cite{others}.  In the simplest case of an
$S^1/Z_2$ orbifold, the matrix representing the action of the $Z_2$
symmetry in field space may not commute with all the generators of the
gauge symmetry.  Boundary conditions may be chosen so that different
components of the gauge multiplet have different $Z_2$ parities,
leaving only some with zero modes after the theory is dimensionally
reduced.  The fact that the zero-mode spectrum includes incomplete
multiplets of the gauge group indicates that the symmetry has been
broken.  Although no Higgs fields are involved, longitudinal gauge
boson scattering amplitudes are well behaved at high
energies~\cite{terning}.  The same approach may be employed to project
away the zero-modes~\cite{ant1} of the color-triplet Higgs in SU(5) GUTS,
naturally resolving the doublet-triplet splitting
problem~\cite{kawamura,others}.

In the simplest orbifold constructions, the orbifold parity commutes
with the diagonal generators of the original gauge symmetry, so that
the unbroken subgroup has the same rank.  For symmetry breakings like
SU(5)$\rightarrow$SU(3)$_C
\times$SU(2)$_W\times$U(1)$_Y$,~\cite{kawamura,others} or
SU(3)$_W\rightarrow$SU(2)$_W\times$U(1)$_Y$~\cite{ant2,weaksu3}, the
breaking by orbifold boundary conditions provides an economical
approach for constructing models.  However, larger groups, like $E_6$
or $E_8$ can only be broken directly to the standard model gauge group
and, at best, a product of additional U(1) factors~\cite{e6e8}.  One
must then rely on the conventional Higgs mechanism to complete the
breaking of the residual GUT symmetry.  In this paper, we will
consider the use of more general boundary conditions to break such
unified symmetries directly to the standard model gauge group, and
hence, to reduce the rank of the original group.  This approach has
been discussed in the context of Higgsless electroweak symmetry
breaking~\cite{terning,morenohiggs}; here we will employ the same
technique at a high scale, while retaining the ordinary Higgs
mechanism for the breaking of electroweak symmetry.  This choice
allows us to eliminate the often complicated and problematic
GUT-breaking Higgs sector, while allowing for the easy generation of
light fermion masses.

The unified theory we consider is based on the `trinified' gauge group
$G_T=$SU(3)$_C\times$SU(3)$_L\times$SU(3)$_R\ltimes
Z_3$~\cite{trin1,trin2,trin3,trin4,trin5,trin6}.  The semidirect
product (indicated by the symbol $\ltimes$) provides for a symmetry
that cyclically permutes the gauge group labels $C$, $L$, and $R$.
Hence, the SU(3)$^3$ representation (rep) $({\bf 1},{\bf 3},{\bf
\bar{3}})$ is part of the trinified rep
\begin{equation}
{\bf 27}=({\bf 1},{\bf 3},{\bf \bar{3}}) \oplus ({\bf \bar{3}},{\bf
1},{\bf 3}) \oplus ({\bf 3},{\bf \bar{3}},{\bf 1}) \,\,\, .
\end{equation}
Moreover, the $Z_3$ symmetry assures the equality of the three SU(3)
gauge couplings at the GUT scale. As originally pointed out in
Ref.~\cite{trin1}, an appropriate embedding of U(1)$_Y$ in
SU(3)$_L\times$SU(3)$_R$ yields the familiar GUT-scale prediction
$\sin^2\theta=3/8$.  We review this construction in
Section~\ref{sec:formalism}.  We will work with a supersymmetric
trinified theory in which the $G_T$ gauge multiplet may propagate in a
single extra dimensional interval.  We first consider the simplest
case in which all the matter and Higgs fields are confined to a brane on
which $G_T$ is broken.  Working in an effective theory of
gauge-symmetry-breaking `spurions' on this brane, we establish the boundary
conditions necessary to break the bulk gauge group to that of the standard
model, $G_T\rightarrow G_{SM}$. We also include the couplings of these spurions
to the matter multiplets of the theory.  In the limit in which the symmetry
breaking parameters are taken to infinity, we obtain the Higgsless limit of
the GUT-breaking sector.  In particular, the mass scale for the
heavy gauge multiplets becomes determined by the compactification radius,
and all exotic matter fields are decoupled from the theory.  The low-energy
theory is simply that of the minimal supersymmetric standard model (MSSM),
with a set of massive gauge multiplets at a scale lower than that of
conventional supersymmetric unification, $2\times 10^{16}$~GeV.  We
show that unification is nonetheless preserved.  Above the compactification
scale, the differential gauge running ({\em i.e.}, $\alpha_i^{-1}(\mu)-
\alpha_j^{-1}(\mu)$ for $i\neq j$) is logarithmic, a feature that has been
noted before in the case of SU(5) GUTS broken on a boundary~\cite{nomura}.  We
then show that viable alternative theories exist in which the Higgs and/or
matter multiplets are allowed to propagate in the bulk space, and we discuss
the boundary conditions on these fields.  In this case, the exotic matter 
fields remain part of the theory, but with large masses set by the 
compactification radius.

Our paper is organized as follows.  In Section~\ref{sec:formalism}, we
review the symmetry breaking in conventional trinification models, and
describe some of the main phenomenological features of these theories.
In Section~\ref{sec:break}, we give the extra-dimensional construction
of supersymmetric SU(3)$^3$, determine the boundary conditions
necessary to break the gauge group down to that of the standard model,
and study the Higgsless limit of the GUT-breaking sector. In
Section~\ref{sec:unify}, we study gauge unification in our minimal
model, while in Section~\ref{sec:chiral} we discuss the possibility of
allowing chiral multiplets in the bulk. In Section~\ref{sec:conc}, we
summarize our conclusions.

\section{Framework}\label{sec:formalism}

Trinification~\cite{trin1,trin2,trin3,trin4,trin5,trin6} is based on the 
gauge group $G_T = $SU(3)$_{C}\times$SU(3)$_{L}\times $SU(3)$_{R}\ltimes Z_3$, 
where $\ltimes$ indicates a semidirect product.  The $Z_3$ symmetry cyclically
permutes the gauge group labels $C$, $L$ and $R$, ensuring a single
unified coupling at the GUT scale. $G_T$ reps consist of the sum of
cyclically permuted SU(3)$^3$ reps.  For example, the gauge fields are
in the ${\bf 24}$-dimensional rep
\begin{equation}
A^\mu_T({\bf 24})=A^\mu_{C}({\bf 8},{\bf 1},{\bf 1})+
A^\mu_{L}({\bf 1},{\bf 8},{\bf 1})+A^\mu_{R}({\bf 1},{\bf 1},{\bf 8}).
\end{equation}
Here, $A_C^\mu$ represent the eight gluon fields of the standard model,
while only some of the $A_L^\mu$ and $A_R^\mu$ above correspond to electroweak
gauge bosons.  The SU(2)$_W$ gauge group of the standard model is
contained entirely in SU(3)$_L$; writing $A = A^a T^a$, then the
SU(2)$_W$ gauge bosons $W^a$ correspond to $A_L^a$ for $a=1\ldots 3$.
On the other hand, the hypercharge gauge boson is a linear combination of
$A_L^8$, $A_R^3$ and $A_R^8$.  The choice
\begin{equation}
A^{\mu}_Y=-\frac{1}{\sqrt{5}}(A^8_L+\sqrt{3}A^3_R+A^8_R)^\mu,
\end{equation}
yields the standard GUT-scale prediction $\sin^{2}\theta_{W}=3/8$.
The pattern of gauge symmetry breaking is achieved via one or more Higgs
fields in the ${\bf 27}$-dimensional rep,
\begin{equation}
\phi({\bf 27}) = \phi({\bf 1},{\bf 3},{\bf \bar{3}}) +
\phi({\bf 3},{\bf \bar{3}},{\bf 1}) + \phi({\bf \bar{3}},{\bf 1},{\bf 3})
\,\,\,.
\end{equation}
Only the first SU(3)$^3$ factor in this rep allows for color-singlet vacuum
expectation values (vevs) that may break $SU(3)^3$ down
to $SU(3)_C \times SU(2)_L \times U(1)_Y$:
\begin{eqnarray}
\phi\mathbf{(1,3,\bar{3})} = \left (
\begin{array}{lll}
 \hat{0} & 0 & 0 \\
0 &  \hat{0} & \hat{0} \\
0 & v_2    & v_1
\end{array}
\right ) \,\,\, .
\end{eqnarray}
Here, $v_i$ represent the GUT-scale vevs, while hatted entries denote
components capable of eventually breaking the electroweak gauge group.
Spontaneous symmetry breaking renders twelve of the original gauge bosons with
masses of order the GUT scale.  Interestingly, these massive gauge
bosons are integrally charged and cannot generate dimension-six operators
that contribute to proton decay. Depending on the number of Higgs multiplets
and their couplings to the matter fields, proton decay may still occur via
color-triplet Higgs exchange.

Standard model fermions are embedded economically in the ${\bf 27}$-dimensional
representation.  In SU(5) language, the ${\bf 27}$ decomposes as
\begin{equation}
{\bf 27}= [{\bf 10} \oplus {\bf \bar{5}}] \oplus {\bf 5} \oplus {\bf \bar{5}} 
\oplus {\bf 1} \oplus {\bf 1} \,\,\, .
\end{equation}
The reps in brackets correspond to a full standard model generation, while
the remaining reps are exotic.  Thus the exotic fields include left- and
right-handed fermions with the quantum numbers of a charge $-1/3$ weak singlet
quark ($B$), a hypercharge $-1/2$ weak doublet lepton ($E^0$, $E^-$) and
an electroweak singlet ($N$).  Using the notation
\begin{eqnarray}
\psi({\bf 27}) & = &
\psi\mathbf{(1,3,\bar{3})}+\psi\mathbf{(\bar{3},1,3)}
+\psi\mathbf{(3,\bar{3},1)}
          \\ & \equiv & \psi_{C}+\psi_{L}+\psi_{R}  \,\,\, ,
\end{eqnarray}
we may choose an SU(2)$_W$ basis in which the fermion reps take the
matrix form
\begin{eqnarray}
\psi_{c} &=& \left (
\begin{array}{lll}
E^{0 c}       & E   & e \\
-E^{c}        & E^0 & \nu   \\
e^{c} & N^{c}   & N
\end{array}
\right ),\;\; \psi_{L} = \left (
\begin{array}{lll}
u^{c}_{\bar{r}}  & u^{c}_{\bar{g}} & u^{c}_{\bar{b}} \\
d^{c}_{\bar{r}}  & d^{c}_{\bar{g}} & d^{c}_{\bar{b}} \\
B^{c}_{\bar{r}}  & B^{c}_{\bar{g}} & B^{c}_{\bar{b}}
\end{array}
\right ),\;\; \psi_{R} = \left (
\begin{array}{lll}
u_r & d_r & B_r \\
u_g & d_g & B_g \\
u_b & d_b & B_b
\end{array}
\right ),
\end{eqnarray}
where all entries are left-handed. In supersymmetric trinification,
these matrices are composed of left-handed chiral superfields, with
each entry indicating the fermionic component.  Yukawa couplings
necessarily involve invariants formed by taking the product of three
${\bf 27}$'s.  These come in two types,
\begin{equation}
Z_{3}[\psi_{R} \psi_{L} \phi_{C i}]\,\,\, , \label{eq:quarkmass}
\end{equation}
and
\begin{equation}
Z_{3}[\psi_{C} \psi_{C} \phi_{C j}] \,\,\, .  \label{eq:lepton}
\end{equation}
We use the symbol $Z_{3}$ to represent the cyclic permutation of
$R$, $L$ and $C$, {\em e.g},
\begin{equation}
Z_{3}[\psi_{R} \psi_{L} \phi_{C}]=\psi_{R} \psi_{L} \phi_{C}+\psi_{C} \psi_{R}
\phi_{L}+\psi_{L} \psi_{C} \phi_{R}.
\end{equation}
The index on the field $\phi_C$ takes into account the possibility that
there may be more than one ${\bf 27}$-plet Higgs field. If there is only one
Higgs ${\bf 27}$, then both the up- and down-type quark Yukawa couplings
for a given generation originate from a single $G_T$-invariant interaction, of
the form shown in Eq.~(\ref{eq:quarkmass}).  This implies the incorrect
GUT-scale mass relation~\cite{trin1}
\begin{equation}
\frac{m_{u}}{m_{d}}=\frac{m_{c}}{m_{s}}=\frac{m_{t}}{m_{b}}. \label{eq:ratio}
\end{equation}
Therefore, at least two Higgs ${\bf 27}$'s must couple to the quarks via
Eq.~(\ref{eq:quarkmass}).  Generally, the same set of Higgs fields will couple
to the leptons via Eq.~(\ref{eq:lepton}) and proton decay may proceed via
color-triplet Higgs exchange.  If a third Higgs ${\bf 27}$-plet is introduced
that couples to the leptons only, then proton decay can be prevented
by imposing a global symmetry on the Higgs sector that prevents mixing
between the third Higgs and the other two.  This, however, leads to
a symmetry-breaking sector that seems somewhat contrived.

It is conventionally assumed that the vevs $v_{1}$ and $v_{2}$
arise in separate Higgs \textbf{27}-plets:
\begin{eqnarray}
\phi\mathbf{(1,3,\bar{3})} &=& \left (
\begin{array}{lll}
0       & 0   & 0 \\
0       & 0   & 0   \\
0       & 0   & v_1
\end{array}
\right ),\;\; \chi\mathbf{(1,3,\bar{3})} = \left (
\begin{array}{lll}
0  & 0   & 0 \\
0  & 0   & 0 \\
0  & v_2 & 0
\end{array}
\right )\,\,\, .
\end{eqnarray}
The superpotential terms responsible for quark and lepton masses can
now be determined from the invariants Eq.(\ref{eq:quarkmass}) and
Eq.(\ref{eq:lepton}),
\begin{equation}
W_Q=(\psi_L)^j_i(\psi_R)^i_k[g_1(\phi_C)^k_j+g_2(\chi_C)^k_j]\,\,\, ,
\end{equation}
\begin{equation}
W_L= \frac{1}{2} h(\psi_C)^i_\alpha(\psi_C)^j_\beta[h_1(\phi_C)^k_\gamma
+h_2(\chi_C)^k_\gamma]
\epsilon_{ijk}\epsilon^{\alpha\beta\gamma} \,\,\, .
\end{equation}
These may be expanded, yielding
\begin{equation}
W =  g_2 v_2\, d^c B + g_1 v_1 B^c B
+ v_1 h_1 \,\epsilon_{ij}  L_H^{c i} L_H^j
- v_2 h_2 \epsilon_{ij}  L_H^{c i} L^j\,\,\, ,
\label{eq:bigmasses}
\end{equation}
where the lepton doublets are defined by  $L_H = (E^0, E)$, $L=(\nu, e)$, and
$L_H^c=(-E^c,E^{c 0})$.  Clearly, one linear combination of $B^c$ and $d^c$, 
and of $L_H$ and $L$, remain unaffected by GUT symmetry 
breaking\footnote{Ref.~\cite{trin2} states that no light lepton eigenstate
will remain if $h_2\neq 0$.  This is not correct, since unbroken electroweak
symmetry assures that a massless eigenstate must remain.}, and should be
identified with the physical right-handed down quark and lepton doublet
superfields:
\begin{eqnarray}
d^c_{phys} & = & (-g_2 v_2 \, B^c + g_1 v_1 \, d^c)/
\sqrt{g_1^2v_1^2 + g_2^2 v_2^2}
\nonumber \\
L_{phys} & = & (h_2 v_2  \, L_H + h_1 v_1  \, L)
/\sqrt{h_1^2 v_1^2 + h_2^2 v_2^2}
\,\,\, .
\end{eqnarray}
The masses of the heavy quark and lepton states remaining in
Eq.~(\ref{eq:bigmasses}) are given by
\begin{equation}
m_{B, B^c_{phys}}=(g_1^2 v_1^2+g_2^2 v_2^2)^{1/2},
\end{equation}
\begin{equation}
m_{L^c_H, L_{H, phys}} = (h_1^2 v_1^2 + h_2^2 v_2^2)^{1/2} \,\,\,.
\end{equation}
For this minimal choice of symmetry breaking, the singlets $N^c$ and
$N$ remain massless. However, as we discuss in the next section, vevs
in other Higgs field representations can give masses to these states
as well.

We will not discuss the structure of the Higgs sector in conventional
trinified theories since our goal is to dispense with this sector
entirely.  We henceforth consider supersymmetric trinified theories
embedded in $4+1$ spacetime dimensions.  As in Ref.~\cite{terning},
we assume that the extra spatial dimension is compact, and runs over
the interval $y=0$ to $y=\pi R$.  We will always assume that the
$G_T$ gauge multiplet propagates in the bulk, and we will consider
consistent boundary conditions that allow us to break this gauge
group directly to that of the standard model upon compactification.
The radius of compactification is a free parameter that we will
determine based on the condition that supersymmetric gauge unification is
preserved. We first consider the simplest case in which all matter
and Higgs fields are placed on the $y=\pi R$ brane, and afterwards
discuss the possibility of placing chiral multiplets in the bulk.

In all cases, we will treat the symmetry breaking on the $\pi R$
brane in an effective theory approach.  We will introduce $G_T$
breaking spurions $\{\Phi_i\}$ on this brane and consider both their couplings
to brane-localized fields, as well as their effect on the 5D wave function
of fields in the bulk.  Historically, the term ``spurion" refers to a
symmetry-breaking parameter that is taken to transform as a spurious
field, so that it may be included consistently in an effective Lagrangian.
In the present case, one may think of the spurions as a collection
of brane Higgs vevs, that can plausibly arise in some ultraviolet
completion.  Since we will focus on the limit in which these vevs
are taken to infinity, we will not defend any particular ultraviolet
theory.  Partial examples will be given only to justify the
consistency of the boundary conditions that we assume. In a few
instances, we will require higher-dimension operators involving
the spurions, which necessarily involve some cut off $\Lambda$.
In the decoupling limit, we will take both $\Phi$ and $\Lambda$ to
infinity in fixed ratio.  In other words, we do not assign $\Lambda$
to some physical scale, but use this limiting procedure to obtain
a consistent Higgsless low-energy effective theory that could otherwise
be defined {\em ab initio}.

\section{Symmetry Breaking}\label{sec:break}

We choose to break the trinified gauge group at the $y=\pi R$ brane. For
a generic gauge field $A^\mu$, the boundary conditions
\begin{equation}
\partial_5 A^\mu(x^\nu,0)=0  \,\,\,\,\,\mbox{ and }\,\,\,\,\,
\partial_5 A^\mu(x^\nu,\pi R)=V A^\mu(x,\pi R)
\label{eq:xdbcs}
\end{equation}
lead to a mode expansion of the form
\begin{equation}
f_k(y) = N_k \cos(M_k y) \,\,\, ,
\end{equation}
where $M_k$ is given by the transcendental equation
\begin{equation}
M_k \tan(M_k \pi R) = - V \,\,\, ,
\label{eq:transcend}
\end{equation}
and where the normalization
\begin{equation}
N_k = \frac{\sqrt{2}}{\sin(M_k \pi R)}[\pi R(1+M_k^2/V^2)-1/V]^{-1/2} \,\,\,
\end{equation}
assures that $\int_0^{\pi R} f^2 = 1$~\cite{terning}. Note that the
symmetry breaking parameter $V$ has dimensions of mass.  The
nontrivial boundary condition in Eq.~(\ref{eq:xdbcs}) can be realized
in an ultraviolet completion of the theory in which a brane localized
Higgs field $\sigma$ is responsible for the symmetry breaking.  The
brane equations of motion for the field $A^\mu$ includes terms
localized at $y= \pi R$ from the start, as well as surface terms
obtained from integrating the bulk action by parts. In particular, the
kinetic terms
\begin{equation}
S_{KE} \supset \int d^4 x \int_0^{\pi R} dy \, [-\frac{1}{2} 
F_{5\nu} F^{5\nu} +
D^\mu \sigma^\dagger D_\mu \sigma \, \delta(y-\pi R)]
\end{equation}
include
\begin{equation}
S_{KE} \supset \int d^4 x \int_0^{\pi R} dy \,
[-\partial_5 A_\nu \partial_5 A^\nu +
\frac{g_5^2}{2} \langle\sigma\rangle^\dagger \langle\sigma\rangle
A^\mu A_\mu \, \delta(y-\pi R)]  \,\,\, .
\end{equation}
Variation of this portion of the action with respect to $A^\nu$ yields
a constraint at $y=\pi R$,
\begin{equation}
-\partial_5 A_\nu + \frac{g_5^2}{2} \langle\sigma\rangle^\dagger
\langle\sigma\rangle A^\nu = 0 \,\,\,,
\end{equation}
which corresponds to the desired boundary condition if one identifies
$V \equiv g_5^2\langle\sigma\rangle^\dagger \langle\sigma\rangle/2$. Since the
5D gauge coupling $g_5$ has mass dimension $-1/2$, one finds that $V$ has
dimensions of mass, as before.

Cs\'{a}ki, Grojean, Murayama, Pilo, and Terning~\cite{terning}, have
demonstrated that the boundary conditions given in Eq.~(\ref{eq:xdbcs})
require a brane-localized Higgs field to cancel contributions to scattering
amplitudes that grow with energy as $E^2$.  However, a remarkable feature
of brane-localized breaking of gauge symmetries is that one can decouple the
Higgs field without decoupling the massive gauge multiplets as well. In the
limit that $\langle\sigma\rangle$, and hence $V$, are taken to infinity, one
finds from Eq.~(\ref{eq:transcend}) that the KK mass spectrum becomes
\begin{equation}
M_n \approx \frac{M_c}{2} (2 n + 1) (1+\frac{M_c}{\pi V} + \cdots) \,\,\, ,
\end{equation}
where $M_c$ is the compactification scale $1/R$.  Thus, the low-energy
theory has no Higgs fields, and the KK tower for the  gauge fields is
shifted by $+M_c/2$ relative to the tower one would obtain if $V$ were
set to zero.

In the case of $G_T$, the first SU(3) factor corresponds to the unbroken
color group, so we may immediately write down the boundary conditions on
the gluon\ fields $A_C^\mu$,
\begin{equation}
\partial_5 A_C^\mu(x,0)=\partial_5 A_C^\mu(x,\pi R)=0 \,\,\, .
\end{equation}
Similarly, an SU(2) subgroup of the second SU(3) factor remains unbroken,
so that
\begin{equation}
\partial_5 A_L^a(x,0) = \partial_5 A_L^a(x,\pi R)= 0 \mbox{ for } a=1\ldots 3
\end{equation}
Since the only remaining unbroken group is a U(1) factor, all gauge fields
corresponding to off-diagonal generators must become massive.  Thus, we
require that
\begin{equation}
\partial_5 A_L^a(x,0) = 0 \mbox{ , } \partial_5 A_L^a(x,\pi R)=
V_L A_L^a(x,\pi R)
\mbox{ for } a=4 \ldots 7 \,\,\, ,
\end{equation}
\begin{equation}
\partial_5 A_R^a(x,0) = 0 \mbox{ , } \partial_5 A_R^a(x,\pi R)=
V_R A_R^a(x,\pi R)
\mbox{ for } a=1,2,4 \ldots 7  \,\,\, .
\end{equation}
The remaining U(1) factors are more interesting.  As we showed in the previous
section, the embedding of hypercharge within
SU(3)$_L\times$SU(3)$_R$ that leads to the prediction $\sin^2\theta=3/8$
requires that the hypercharge gauge boson be identified with the linear
combination
\begin{equation}
A_Y^\mu = -\frac{1}{\sqrt{5}}(A^8_{L}+\sqrt{3} A^3_R + A^8_R)^\mu
\label{eq:hyp}
\end{equation}
Thinking in terms of an ultraviolet completion, suitable Higgs fields must
generate a brane gauge boson mass matrix with a zero eigenvalue corresponding
to the eigenvector $(-1/\sqrt{5},-\sqrt{3}/\sqrt{5},-1/\sqrt{5})$.  The
only other necessary constraint on this matrix is that the remaining
eigenvalues must be non-vanishing. Restricting ourselves to real entries, for
the sake of simplicity, we may parameterize the remaining boundary conditions
as follows:
\begin{equation}
\partial_5 A_L^8(x,0)= \partial_5 A^3_R(x,0)= \partial_5 A^8_R(x,0)=0 \,\,\, ,
\end{equation}
\begin{equation}
\partial_5 \left[\begin{array}{c} A_L^8(x,\pi R) \\ A_R^3(x,\pi R) \\
A_R^8(x,\pi R)
\end{array}\right] = \left(\begin{array}{ccc}
V_1 & -\frac{1}{2\sqrt{3}}(V_1+V_3) & -\frac{1}{2}(V_1- V_3) \\
-\frac{1}{2\sqrt{3}}(V_1+V_3)  & \frac{1}{6}(V_2+V_3) &
\frac{1}{2\sqrt{3}} (V_1-V_2) \\
-\frac{1}{2}(V_1- V_3) & \frac{1}{2\sqrt{3}}(V_1-V_2) & \frac{1}{2} (V_2 - V_3)
\end{array}\right)\left[\begin{array}{c} A_L^8(x,\pi R) \\ A_R^3(x,\pi R)
\\ A_R^8(x,\pi R)
\end{array}\right]
\label{eq:thematrix}
\end{equation}
Finally, we consider the $A^5$ components.  In a nonsupersymmetric theory,
we could impose the boundary conditions $A^5(x,0)=A^5(x,\pi R)=0$ on all
the gauge fields so that no additional light scalar states remain in the 
4D theory.  In the supersymmetric case, $A^\mu$ and $A^5$ live within a
vector $V$ and chiral $\Phi_V$ superfield, respectively.  Since 
supersymmetry is unbroken, the fermionic components of $V$ and $\Phi_V$
(say, $\lambda$ and $\psi$) must form  Dirac spinors  with the same mass 
spectrum as the gauge fields~\cite{nomura}.  Since these masses originate 
from terms of the form $\partial_5 \lambda \psi$, the 5D wave function 
of $\Phi_V$ must be proportional to $\sin M_k y$, with $M_k$ given as before.

If one were to assume an ultraviolet completion involving only the minimal
Higgs content of conventional 4D trinified theories (localized on the $\pi R$
brane) one would find that
\begin{eqnarray}
V_1 &=& \frac{2}{3}(v_1^2+v_2^2) g_5^2 \nonumber \\
V_2 &=& \frac{1}{3}(2 \, v_1^2 + 5\, v_2^2) g_5^2 \nonumber \\
V_3 &=& -\frac{1}{3}(2\, v_1^2-4\,  v_2^2)g_5^2  \,\,\, .
\label{eq:inparticular}
\end{eqnarray}
The more general values of the parameters $V_i$ may be thought of as
arising in some arbitrarily complicated GUT-breaking Higgs sector, which
decouples as one takes $V_L$, $V_R$ and $V_i \rightarrow \infty$.  However,
we will not wed ourselves to any particular interpretation of the physics
responsible for generating the symmetry-breaking parameters on the
boundary.

We will proceed with an effective field theory analysis of the
possible symmetry breaking on the $\pi R$ brane. We will introduce the
symmetry breaking systematically in terms of constant spurion fields that
we may treat as transforming in irreducible reps of SU(3)$^3$.  When we
obtain operators that are nonrenormalizable, we will introduce powers of a
cutoff, $\Lambda$ to obtain the proper mass dimension, as discussed
at the end of Section~\ref{sec:formalism}.

A given spurion representation may contribute to the symmetry breaking
parameterized by Eq.~(\ref{eq:thematrix}) provided that it contains standard
model singlet components, with hypercharge defined as in Eq.~(\ref{eq:hyp}),
that develop vevs.  We know immediately of one possibility from the minimal 4D
trinified theory, namely a {\bf 27} with vevs in the
$({\bf 1},{\bf 3},{\bf \bar 3})$ component,
\begin{equation}
\Phi({\bf 1},{\bf 3},{\bf \bar 3}) \sim \left(\begin{array}{ccc}
0 & 0 & 0 \\ 0 & 0 & 0 \\ 0 & v_2 & v_1 \end{array}\right) \,\,\,.
\end{equation}
As described in Section~\ref{sec:formalism}, these vevs give mass to
the heavy fields $B$, $B^c$, $L_H$ and $L_H^c$ while contributing to
the boundary condition on the gauge fields via
Eq.~(\ref{eq:inparticular}).  This rep, however, does not contribute
to the mass of the new singlet leptons, $N^c$ and $N$. Since we wish
to retain only the particle content of the MSSM at the electroweak
scale, we will be more general.  The set of SU(3)$^3$ representations
that appear in the product of two {\bf 27}'s and that are color
singlet are
$({\bf 1},{\bf 3},{\bf \bar 3})$, $({\bf 1},{\bf \bar 6},{\bf 3})$,
$({\bf 1},{\bf 3},{\bf 6})$, and $({\bf 1},{\bf \bar 6},{\bf 6})$.  For each,
we may isolate the components that are SU(2)$_W\times$U(1)$_Y$ singlets. The
results are shown in Table~\ref{table:reps}.
\begin{table}
\begin{center}
\caption{SU(3)$^3$ reps in the product of two trinified {\bf 27}-plets
containing Standard Model singlet components, with hypercharge defined as
in Eq.~(\ref{eq:hyp}). Parentheses delimit indices that are symmetric.}
\label{table:reps}
\vspace{1em}
\begin{tabular}{ccc}\hline\hline
SU(3)$^3$ rep  \quad\quad &  SU(3)$_L\times$SU(3)$_R$ tensor  &
SM singlet components \\
\hline
$({\bf 1},{\bf 3},{\bf \bar 3})$ & $\Phi^a_\alpha$ & $a=3$, $\alpha=2,3$ \\
$({\bf 1},{\bf \bar 6},{\bf \bar 3})$ & $\Phi_{(ab)\alpha}$ & $a=b=3$,
$\alpha=1$\\
$({\bf 1},{\bf 3},{\bf 6})$      & $\Phi^{a(\alpha\beta)}$ & $a=3$,
$(\alpha\beta)=(12),(13)$ \\
$({\bf 1},{\bf \bar 6},{\bf 6})$ & $\Phi_{(ab)}^{(\alpha\beta)}$ &
$(ab)=(33)$, $(\alpha\beta)=(22),(23),(33)$ \\
\hline\hline
\end{tabular}
\end{center}
\end{table}
While the reps $({\bf 1},{\bf \bar 6},{\bf 3})$ and $({\bf 1},{\bf 3},{\bf 6})$
contain standard model singlet components, it turns out that these do not
split the {\bf 27} matter multiplets.  For example, the coupling of the
$({\bf 1},{\bf \bar 6},{\bf \bar 3})$ to two {\bf 27} matter superfields
may be written
\begin{equation}
W = \Psi^a_\alpha \Psi^b_\beta \Phi^{({\bf 1},{\bf \bar 6},{\bf \bar 3})
}_{ab,\gamma} \epsilon^{\alpha\beta\gamma}
\end{equation}
which vanishes for $a=b=3$ and $\gamma=1$, because of the antisymmetry of the
SU(3)$_R$ epsilon tensor.  Of the three new spurion reps in
Table~\ref{table:reps}, only the $({\bf 1},{\bf \bar 6},{\bf 6})$ gives us
something new,
\begin{eqnarray}
W &=& \Psi^a_\alpha \Psi^b_\beta \Phi^{\alpha\beta}_{ab} \nonumber \\
  &=& v_{22} N_R^2 +2 v_{23} N_R N_L + v_{33} N_L^2 \,\,\, .
\label{eq:Nmasses}
\end{eqnarray}
Here $v_{ij}$ corresponds to vevs for the standard model singlet components
of the $({\bf 1},{\bf \bar 6},{\bf 6})$ spurion, as given in
Table~\ref{table:reps}. Hence, we arrive at Majorana and Dirac masses for the
exotic neutral leptons, which may be decoupled from the theory if the $v_{ij}$
are taken to infinity.  Thus we reach the following conclusion:

Gauge symmetry breaking spurions localized at the $\pi R$ brane in the
${\bf 27}$ and ${\bf 108}$ irreducible reps of the trinification group, and
with nonvanishing standard model singlet entries in their
$({\bf 1},{\bf 3},{\bf \bar 3})$ and $({\bf 1},{\bf \bar 6},{\bf 6})$
components, respectively, break the trinification gauge group down to the
standard model, and yield the MSSM matter content at low energies.  In
the limit that all the symmetry breaking parameters are taken to infinity,
we obtain Higgsless trinification breaking with an incomplete matter
multiplet located at the $\pi R$ brane.

This picture is pleasing since any physics on the brane associated with an
ultraviolet completion that might lead to proton decay has been decoupled
away.  The only issue we have not taken into account is the mechanism for
breaking electroweak symmetry and the generation of light fermion masses.
We may easily incorporate the standard Higgs mechanism for electroweak
symmetry breaking by introducing ${\bf 27}$ and ${\bf \overline{27}}$ Higgs
superfields on the $\pi R$ brane, $\Psi_H$ and $\Psi_{\bar{H}}$, respectively.
(These are distinguished from matter superfields by unbroken matter or 
R-parity, which we assume throughout.)  We identify $H$ ($\bar{H}$) as the 
doublet Higgs field with hypercharge $1/2$ ($-1/2$) living inside the 
multiplet  $\Psi_H$ ($\Psi_{\bar{H}}$). We also introduce another spurion 
rep, the ${\bf 192}$, which includes the color singlet rep
$\Omega \sim ({\bf 1},{\bf 8},{\bf 8})$.  Assuming that the nonvanishing,
standard model singlet components of $\Omega$ are given by
\begin{equation}
\Omega^{\alpha\beta}_{ab} = v_\Omega\, {T^8}^a_b \, {T^3}^\alpha_\beta
\end{equation}
then the couplings
\begin{equation}
W = H^a_\alpha (\mu \, \delta^\alpha_\beta \delta^b_a
+h\, \Omega^{b\beta}_{a\alpha}) \bar H_b^\beta
\end{equation}
will provide high-scale $\mu$ terms for all members of the Higgs multiplet,
except for the weak doublets $H$ and $\bar{H}$, providing that
$\mu=-4\,\sqrt{3} \,h \,v_\Omega$.  Thus, in this approach, we simply impose
a fine-tuning of the parameters to arrange for a doublet-triplet
splitting~\footnote{Higher order combinations of the other spurions
may generate a $({\bf 1},{\bf 8},{\bf 8})$; we assume a fine tuning of the
sum of all such contributions.}.  However, since we ultimately take the 
limit in which $v_\Omega \rightarrow \infty$, as with the other 
symmetry-breaking spurions, there is no sign of this fine-tuning in the 
low-energy theory.  From a low-energy perspective, it is completely 
consistent to assign two electroweak Higgs doublets to the brane in the 
GUT-Higgsless limit.

One feature of this solution that needs clarification is the coupling of
these Higgs doublets to the matter fields.  While the up-quark Higgs fields $H$
lives in a ${\bf 27}$ and couples to the matter fields via the conventional
cubic interactions of 4D trinified theories, the down-type Higgs fields
$\bar{H}$ lies in a ${\bf \overline{27}}$ and does not couple directly.
Nonetheless,we may arrange for a suitable down quark Yukawa matrix by
introducing a ${\bf \overline{27}}$ spurion with the same nonvanishing
components as the ${\bf 27}$ spurion that we have already considered.  Then
the down quark Yukawa matrix will originate via a higher-dimension operator
\begin{equation}
\frac{1}{\Lambda} Z_3 [ \Phi({\bf 1},{\bf \bar 3},{\bf 3})
\bar H ({\bf 1},{\bf \bar 3},{\bf 3})
\Psi({\bf \bar 3},{\bf 1},{\bf 3}) \Psi({\bf 3},{\bf \bar 3},{\bf 1})] \,\,\,.
\end{equation}
We may generate the down quark Yukawa couplings by fixing the ratio of
the spurion vev to $\Lambda$, and taking both to infinity in the Higgsless
limit.

\section{Gauge Unification}\label{sec:unify}

By breaking the GUT gauge group through boundary conditions, the heavy vector
superfields that have GUT-scale masses in 4D trinified theories instead
have zero-modes with mass $M_c/2$ in the exact Higgsless limit.  The
SU(3)$_C\times$SU(2)$_W\times$U(1)$_Y$ quantum numbers of these states are
given by
\begin{equation}
V_H \sim ({\bf 1},{\bf 2},1/2) \oplus ({\bf 1},{\bf 1},1)
\oplus ({\bf 1},{\bf 1},1) \oplus ({\bf 1},{\bf 1},0) \oplus 
({\bf 1},{\bf 1},0)
\,\,\, ,
\end{equation}
where the hypercharges are shown here with their standard, rather than
their GUT, normalization.  KK modes of the ordinary MSSM vector
superfields begin at $M_c$.  The two towers of massive states are thus
uniformly shifted with respect to each other by $M_c/2$.  Each KK
level in these towers consists of an $N=2$ supersymmetric multiplet,
which includes both a vector and a chiral superfield.  The beta
function contributions from these towers are indicated in
Table~\ref{betatable}.  Notice that the sum of all the KK gauge
multiplet contributions to the beta functions is $(-6,-6,-6)$; if the
two massive towers were degenerate level by level, they would affect
gauge coupling running universally and have no effect on the quality
of unification. However, the $M_c/2$ splitting separates these states
into two subsets, each contributing nonuniverally to the beta
functions. The shifted towers therefore provide a large number of
threshold corrections to the differential gauge coupling running
$\alpha_i^{-1}(\mu)- \alpha_j^{-1}(\mu)$. There is no reason {\em a
priori} to assume that these corrections will preserve gauge
unification.  In our trinified theory, we will see that they do.
\begin{table}
\begin{center}
\begin{tabular}{ccccc}\hline\hline
 &$(b_1,b_2,b_3)$ \qquad &  \qquad
($\tilde{b_1},\tilde{b_2},\tilde{b_3}$) \\ \hline
$(V,\Phi)_{321} $  &  (0,-6,-9) \qquad & \qquad (0,-4,-6) \\
$(V,\Phi)_{heavy}$       &   - \qquad & \qquad (-6,-2,0)
\\ $H$, $\overline{H}$ & ($\frac{3}{5}$,1,0) \qquad &\qquad  -  \\
Matter & (6,6,6) \qquad & \qquad  -  \\ \hline Total &
($\frac{33}{5}$,1,-3) \qquad & \qquad (-6,-6,-6)  \\ \hline\hline
\end{tabular}
\end{center}
\caption{Contributions to the beta function coefficients from
the zero modes ($b_i$) and the KK levels ($\tilde{b}_i$) in
our minimal scenario. Here $\Phi$ represents a chiral
multiplet in the adjoint rep.}\label{betatable}
\end{table}

\begin{figure}[t]
\epsfxsize 3.3 in \epsfbox{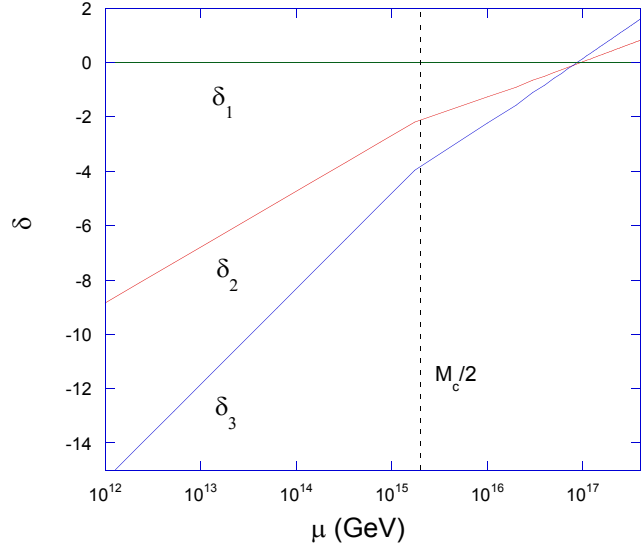} \caption{Gauge unification for
$M_c=4\times 10^{15}$~GeV.}
\label{fig:one}
\end{figure}

While the individual $\alpha_i^{-1}$ experience power-law running above
$M_c/2$, a remarkable feature of this tower of threshold corrections is
that the $\alpha_i^{-1}(\mu)- \alpha_j^{-1}(\mu)$ evolve logarithmically.
This behavior was pointed out by Nomura, Smith and Weiner~\cite{nomura}
in the context of a supersymmetric SU(5) GUT broken on a brane. Thus,
theories of this type unify logarithmically, in contrast
to the first examples of higher-dimensional gauge unification discussed 
in Refs.~\cite{ddgplus}.  For our analysis, we follow the conventions of 
Ref.~\cite{nomura}:  We first define gauge coupling differences with 
respect to $\alpha_1^{-1}$,
\begin{equation}
\delta_i(\mu)=\alpha^{-1}_i(\mu)-\alpha^{-1}_1(\mu). \label{eq:delta}
\end{equation}
Unification occurs when $\delta_2=\delta_3=0$.  Above $M_c/2$,
Eq.(\ref{eq:delta}) can be written as
\begin{equation}
\delta_i(\mu)=\delta_i(M_c/2)-\frac{1}{2\pi}R_i(\mu)
\label{eq:dieq}
\end{equation}
where $R_i(\mu)$ represents the differential logarithmic running between
all the thresholds from $M_c/2$ up the the renormalization scale $\mu$.
For trinified gauge multiplets in the bulk only, we find
\begin{equation}
R_2(\mu)=-\frac{28}{5} \log ( \frac{\mu}{M_c/2})-4\sum_{0<nM_c<\mu}\log(
\frac{\mu}{nM_c})+4\!\!\sum_{0<(n+1/2)M_c<\mu}\!\!\log(\frac{\mu}{[n+1/2]M_c}),
\end{equation}
\begin{equation}
R_3(\mu)=-\frac{48}{5} \log ( \frac{\mu}{M_c/2})-6\sum_{0<nM_c<\mu}\log(
\frac{\mu}{nM_c})+6\!\!\sum_{0<(n+1/2)M_c<\mu}\!\!\log(\frac{\mu}{[n+1/2]M_c}).
\end{equation}
If the two towers of massive modes were degenerate, the last two terms in
each of the equations above would have exactly canceled, and the
$R_i$'s would be the same as in the MSSM.  The overall effect of the
threshold corrections is to delay unification, as shown in Fig.~\ref{fig:one}.

The shallower slopes above $M_c/2$ in Fig.~\ref{fig:one} can be
understood by rewriting Eq.~(\ref{eq:dieq}) in the form.
\begin{equation}
\delta_i(\mu) = \delta_i(M_c/2)-\frac{1}{2\pi} \delta b_i
\log(\frac{\mu} {M_c/2}) -\frac{1}{2\pi} \Delta \tilde{b}_{321} \!\!
\sum_{n M_c/2 <\mu} (-1)^n \, \log (\frac{\mu}{n M_c/2})
\end{equation}
where we have used the fact that the difference in KK gauge multiplet
beta functions $\Delta \tilde{b}_{321} = - \Delta \tilde{b}_{heavy}$.  The
first and second terms are negative and positive, respectively, and
cancel in the MSSM at the unification point. The new term has positive
coefficient $-\frac{1}{2\pi} \Delta \tilde{b}_{321}$.  However, one
may estimate the sum via integration, and one finds it is well
approximated by $-(1+\log(\mu/M_c))/2$. Thus, the new threshold
corrections serve to reduce the effect of the second term (the MSSM
differential logarithmic running) so that unification is delayed.

\begin{figure}[t]
\epsfxsize 3.3 in \epsfbox{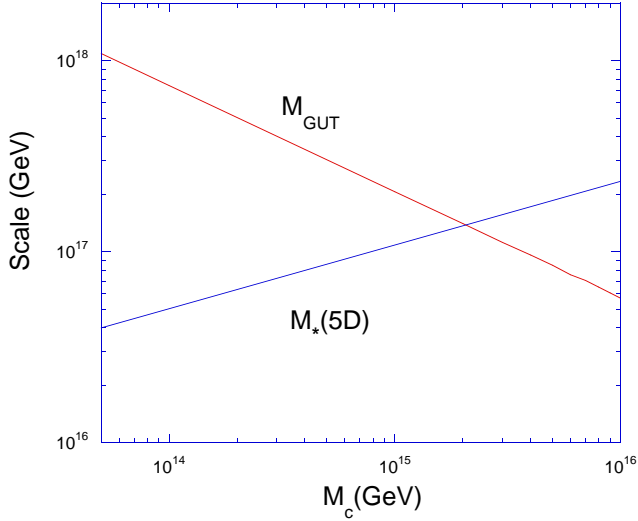} \caption{Unification and
5D Planck scales as functions of $M_c$.  For definitions of scales, see 
the text.}
\label{fig:two}
\end{figure}

In the Higgsless limit, there are two significant physical scales in
the theory: the compactification scale $1/R$, which determines the
masses of the super-heavy states in the theory, and the 5D Planck scale, 
$M_*(5D)$, which determines where gravity becomes important.  In 
Fig.~\ref{fig:two}, we show both the unification scale $M_{{\rm GUT}}$, 
defined as the point at which $\alpha^{-1}_1= \alpha^{-1}_2$, and
$M_*(5D)$, as a function of the compactification scale $M_c$.  These
scales are identical when $M_c \sim 2 \times 10^{15}$~GeV.  For larger
$M_c$, the 5D Planck scale is higher; in this case, one could introduce
other, purely gravitational extra dimensions that again bring the 
higher-dimensional Planck scale in coincidence with $M_{{\rm GUT}}$.
For $M_c \alt 2 \times 10^{15}$~GeV, $M_*(5D)$ is lower than $M_{{\rm GUT}}$
and a field theoretic calculation of gauge coupling unification can no
longer be trusted.  For all values of $M_c$ larger than 
$2 \times 10^{15}$~GeV, the unification scale is increased relative to
that of the 4D MSSM, {\em i.e.}, $2\times 10^{16}$~GeV.  At its maximum
value, $1.4 \times 10^{17}$~GeV, the accuracy of gauge unification
is $\sim 1\%$.  This estimate assumes that brane-localized,
higher-dimension kinetic energy operators have a negligible effect on
the equality of the gauge couplings at the unification scale.  Such an
assumption is reasonable since these effects are volume suppressed by
a factor of $\sim \pi M_*(5D)/M_c$~\cite{nomura}, which is generally
large. Of course, the precise values of the operator coefficients are
unknown, and one cannot rule out the possibility that such operators
are simply not present in the theory.

\section{Other Possibilities} \label{sec:chiral}

In the previous sections, we have allowed all exotic chiral
superfields to be perfectly decoupled in the Higgsless limit.  This
was accomplished by restricting matter and Higgs multiplets to the
$y=\pi R$ brane, and including the most general set of couplings to
the symmetry breaking parameters. In this section, we discuss the
alternative possibility that some (or all) of the ${\bf 27}$'s
propagate in the bulk, along with the gauge multiplets.  Assuming the
same set of symmetry-breaking parameters on the $\pi R$ brane, exotic
fields now acquire masses of order the compactification scale, leaving
the MSSM at low energies.

In general, a bulk matter field consists of an $N=2$ hypermultiplet
$\Psi=(\psi, \psi^c)$, where $\psi$ and $\psi^c$ are each left-handed, 4D $N=1$
chiral superfields; in our case, these fields transform as a ${\bf 27}$ and
a ${\bf \overline{27}}$, respectively. We wish to argue that it is consistent
within our framework to apply the following simple boundary conditions to
elements of the ${\bf 27}$ (and conjugate elements in the
${\bf \overline{27}}$) that we require to become massive:
\begin{equation}
\partial_5 \phi\,|_{y=0}=\phi^c\,|_{y=0} = \phi\,|_{y=\pi R} =
\partial_5 \phi^c\,|_{y=\pi R} = 0 \,\,\,.
\label{eq:morebcs1}
\end{equation}
Here, $\phi$ and $\phi^{c}$ represented the scalar components of
$\Psi$ and $\Psi^c$, respectively. These boundary conditions are satisfied for
\begin{eqnarray}
\phi &=& \sum_k N_k \cos(M_k y) \, \phi^{(k)} \nonumber \\
\phi^c &=& \sum_k N_k \sin(M_k y)\, {\phi^c}^{(k)} \,\,\, ,
\label{eq:modesagain}
\end{eqnarray}
where $N_k = (\pi R/2)^{-1/2}$, and $M_k=(k+1/2)\,M_c$, for
integer $k$. Of course, Eq.~(\ref{eq:modesagain}) solve the bulk equations
of motion $\partial_M\partial^M \phi =0$ provided that the KK modes satisfy
the on-shell relation $p_k^2=M_k^2$.  Since supersymmetry is unbroken, the
same conditions apply to the fermionic components as well.

To show that these boundary conditions are consistent, let us consider
one possible ultraviolet completion.  First, let us generalize
our boundary conditions to
\begin{eqnarray}
\partial_5 \phi\,|_{y=0}=0  &\,\,\,,\,\,\,& \phi^c\,|_{y=0} = 0  \nonumber \\
(-\sin\eta \,\phi^c + \cos\eta \,\phi)\,|_{y=\pi R} = 0 &\,\,\,,\,\,\,&
\partial_5(\cos\eta \,\phi^c + \sin\eta \,\phi)\,|_{y=\pi R} = 0
\label{eq:morebcs}
\end{eqnarray}
which are satisfied by Eq.~(\ref{eq:modesagain}), if
\begin{equation}
\tan(M_k \pi R) = \cot\eta \,\,\, .
\end{equation}
Notice that one linear combination of the fields in Eq.~(\ref{eq:morebcs})
satisfies Dirichlet boundary conditions at $y=\pi R$, while the orthogonal
satisfies has Neumann boundary conditions. The precise linear combination is
determined by the mixing angle $\eta$, which is a free parameter.  Our
desired boundary conditions are obtained from Eq.~(\ref{eq:morebcs}) in the
limit that $\eta \rightarrow 0$.

Now consider the following 5D Lagrangian, with a brane-localized $\mu$-term
\begin{equation}
{\cal L}_5 = \int d^4\, \theta [\psi^\dagger \psi + {\psi^c}^\dagger \psi^c]
+\int d^2\theta \, [\psi^c \partial_5 \psi + \frac{1}{2} \cot\eta \, \psi^2
\delta(y-\pi R)]
\label{eq:braneL}
\end{equation}
Here we have displayed the effective $N=1$ supersymmetric Lagrangian,
following the construction described in Ref.~\cite{nima}.  Extracting the 
purely scalar components, one finds
\begin{eqnarray}
{\cal L}_5&=& F^\dagger F + \partial_\mu \phi^\dagger \partial^\mu \phi
+ {F^c}^\dagger F^c + \partial_\mu {\phi^c}^\dagger \partial^\mu \phi^c
\nonumber \\
&+& [\,\phi^c \partial_5 F +F^c \partial_5 \phi + \cot\eta \,
\phi F \,\delta(y-\pi R)
+ \mbox{ h.c.}]
\end{eqnarray}
Aside from the bulk equations of motion for the auxiliary fields
$F=\partial_5 {\phi^c}^\dagger$ and $F^c=-\partial_5 \phi^\dagger$, one finds
from the nonvanishing surface terms the boundary condition
\begin{equation}
-\phi^c \delta F |_{y=0} + (\phi^c + \cot\eta\, \phi) \delta F |_{y=\pi R}=0
\end{equation}
which is clearly satisfied by the boundary conditions in
Eq~(\ref{eq:morebcs}).  Substituting out the auxiliary fields, one is
left with the Lagrangian
\begin{eqnarray}
{\cal L} &=& \partial_5 {\phi^c}^\dagger \partial_5 \phi^c
-\partial_5^2 {\phi^c}^\dagger \phi^c  + \phi_c^\dagger \partial_5^2\phi_c
+\partial_\mu {\phi^c}^\dagger \partial^\mu \phi^c \nonumber \\
&+&\partial_\mu \phi^\dagger \partial^\mu \phi- \partial_5 \phi^\dagger
\partial_5 \phi-\cot\eta\, \delta(y-\pi R) (\partial_5{\phi^c}^\dagger \phi
+\phi^\dagger \partial_5 \phi^c) \,\,\, .
\end{eqnarray}
Variation of the action with respect to $\phi$ leads to the further brane
constraint
\begin{equation}
 \partial_5 \phi^\dagger \delta\phi |_{y=0}+
\partial_5 (- \phi^\dagger+\cot\eta \,
{\phi^c}^\dagger)\,\delta\phi|_{y=\pi R}=0
\end{equation}
which is satisfied by the remaining boundary conditions in
Eq~(\ref{eq:morebcs}).  Thus, our more general set of  boundary conditions
are consistent with this explicit brane Lagrangian.  In particular, the
simpler boundary conditions in Eq.~(\ref{eq:morebcs1}) arise in the limit
that the coupling $\cot\eta$ is allowed to become nonperturbatively large.

In the context of our previous discussion, the dimensionless brane coupling
proportional to $\cot\theta$ arises at some order in the symmetry breaking
spurions $\Phi$.  Generically,
\begin{equation}
W =-\frac{1}{2} \lambda (\Phi/\Lambda)\, \psi \psi \,\delta(y-\pi R) \,\,\, ,
\end{equation}
where, $\lambda$ is a dimensionless coupling, and $\cot\eta$ is identified
with $\lambda\Phi/\Lambda$.  Any exotic field that decoupled in our earlier
construction, will receive a brane coupling proportional to $\cot\eta$ in
the present one.  Thus, in the $\eta \rightarrow 0$ limit, we recover the
boundary conditions of Eq.~(\ref{eq:morebcs1}) applied to that particular
field, whose zero mode obtains a mass of  $M_c/2$.

If we take this completion literally, then we would want to restrict
$\cot\eta$ by the condition that the coupling $\lambda$ remain
perturbative.  However, we are not wedding ourselves to any particular
origin for the boundary conditions. We will take the example just discussed
as motivation for the consistency of Eq.~(\ref{eq:morebcs1}),
and work in the exact $\eta=0$ limit.  The reader who disagrees with this
approach may simply consider our results an approximation to the explicit
ultraviolet completion discussed above when $\cot\eta$ is taken to be
somewhat strongly coupled.

In the case where the bulk ${\bf 27}$'s are the three standard model
generations, the exotic $N$, $E$ and $B$ fields will become massive
given our choice of brane spurions. Our results for gauge unification will
not be affected since these fields form the complete SU(5) reps
${\bf 5}\oplus{\bf \bar 5}\oplus{\bf 1}\oplus{\bf 1}$.  Another possibility 
is to place the ${\bf 27}$ and ${\bf \overline{27}}$ Higgs multiplets in the 
bulk.  In this case, we have a tower of KK modes beginning at $M_C/2$ for the 
massive components, and a tower beginning at $M_C$ for those components with
massless zero modes.  This leads to an additional threshold correction of
the type discussed in Section~IV.  We find that this tends to spoil
unification for values of $1/R$ that are significantly smaller than
the conventional supersymmetric unification scale, $M_u=2 \times 10^{16}$~GeV.
Thus, this possibility may be realized if $1/R$ and $M_u$ are within a factor
of a few of each other so that unification is preserved to good approximation.

\section{Conclusions}\label{sec:conc}
The breaking of gauge symmetries through the choice of consistent
boundary conditions on an extra dimensional interval provides a
powerful new tool for model building.  Unlike the orbifold case, a
more general choice of boundary conditions allows one to reduce the
rank of the bulk gauge group.  Aside from the breaking of electroweak
symmetry~\cite{terning,others}, this approach is naturally of interest
in the breaking of grand unified and other gauge extensions of the
standard model that have gauge groups with rank greater than four.  We
have demonstrated this explicitly in the case of gauge trinification.
We obtained boundary conditions necessary to break the trinified gauge
group directly down to that of the standard model, while preserving
the GUT-scale relation $\sin^{2}\theta_{W}=3/8$.  Symmetry
breaking was introduced consistently in terms of spurions localized on
the $\pi R$ brane.  In the Higgsless limit, in which these spurions
are taken to infinity, the massive gauge multiplets have zero-modes at
$M_c/2$, where $M_c$ is the compactification scale. In the same limit,
all exotic matter and Higgs fields are decoupled from the theory, and
Higgs-mediated proton decay is avoided.  We retain the light Higgs
doublets of the MSSM, so that light fermion masses may be easily
obtained.  By placing the gauge multiplets in the bulk, there is power
law running due to the KK modes. As in other 5D unified theories with
gauge symmetries broken on a boundary~\cite{nomura}, we find that the
running of the differences $\alpha^{-1}_i-\alpha^{-1}_j$ remains
logarithmic.  For the massive gauge fields in our trinified theory, we
find that unification is preserved, and that the scale at which the
couplings unify is increased.  For $M_c \sim 2 \times 10^{15}$~GeV,
the gauge couplings unify at the 5D Planck mass $1.4 \times 10^{17}$~GeV,
with a percent accuracy at the one-loop level.


\begin{acknowledgments}
J.C. thanks Y.~Nomura for a useful clarification of Ref.~\cite{nomura}.
We thank the NSF for support under Grant No.~PHY-0140012 and supplement
PHY-0352413. In addition, CDC thanks the William and Mary Endowment
Association for its support.
\end{acknowledgments}


\end{document}